\documentclass[a4paper,UKenglish,cleveref, autoref, thm-restate]{lipics-v2021}
\usepackage{fancyvrb}
\usepackage[utf8]{inputenc}

\nolinenumbers

\newsavebox\myv

\newcommand{\SA}{\ensuremath{\mathrm{SA}}}
\newcommand{\LCP}{\ensuremath{\mathrm{LCP}}}
\newcommand{\BWT}{\ensuremath{\mathrm{BWT}}}
\newcommand{\rank}{\ensuremath{\mathrm{rank}}}

\newcommand{\PSV}{\ensuremath{\mathrm{PSV}}}
\newcommand{\NSV}{\ensuremath{\mathrm{NSV}}}

\title{Taxonomic classification with maximal exact matches in KATKA kernels and minimizer digests}
\titlerunning{Taxonomic classification with MEMs}

\author{Dominika Draesslerov\'a}{Czech Technical University in Prague, Czech Republic}{}{}{}
\author{Omar Ahmed}{Johns Hopkins University, USA}{}{}{}
\author{Travis Gagie}{CeBiB \& Dalhousie University, Canada}{}{}{}
\author{Jan Holub}{Czech Technical University in Prague, Czech Republic}{}{}{}
\author{Ben Langmead}{Johns Hopkins University, USA}{}{}{}
\author{Giovanni Manzini}{University of Pisa, Italy}{}{}{}
\author{Gonzalo Navarro}{CeBiB \& University of Chile, Chile}{}{}{}
\authorrunning{D. Draesslerov\'a et al.}

\Copyright{Dominika Draesslerov\'a, Omar Ahmed, Travis Gagie, Jan Holub, Ben Langmead, Giovanni Manzini and Gonzalo Navarro}

\ccsdesc[500]{Theory of computation~Pattern matching}

\keywords{Taxonomic classification, metagenomics, KATKA, maximal exact matches, string kernels, minimizer digests}

\EventEditors{John Q. Open and Joan R. Access}
\EventNoEds{2}
\EventLongTitle{42nd Conference on Very Important Topics (CVIT 2016)}
\EventShortTitle{CVIT 2016}
\EventAcronym{CVIT}
\EventYear{2016}
\EventDate{December 24--27, 2016}
\EventLocation{Little Whinging, United Kingdom}
\EventLogo{}
\SeriesVolume{42}
\ArticleNo{23}
\begin{document}

\maketitle

\begin{abstract}
For taxonomic classification, we are asked to index the genomes in a phylogenetic tree such that later, given a DNA read, we can quickly choose a small subtree likely to contain the genome from which that read was drawn.  Although popular classifiers such as Kraken use $k$-mers, recent research indicates that using maximal exact matches (MEMs) can lead to better classifications.  For example, we can
\begin{itemize}
\item build an augmented FM-index over the the genomes in the tree concatenated in left-to-right order;
\item for each MEM in a read, find the interval in the suffix array containing the starting positions of that MEM's occurrences in those genomes;
\item find the minimum and maximum values stored in that interval;
\item take the lowest common ancestor (LCA) of the genomes containing the characters at those positions.
\end{itemize}
This solution is practical, however, only when the total size of the genomes in the tree is fairly small.  In this paper we consider applying the same solution to three lossily compressed representations of the genomes' concatenation:
\begin{itemize}
\item a KATKA kernel, which discards characters that are not in the first or last occurrence of any $k_{\max}$-tuple, for a parameter $k_{\max}$;
\item a minimizer digest;
\item a KATKA kernel of a minimizer digest.
\end{itemize}
With a test dataset and these three representations of it, simulated reads and various parameter settings, we checked how many reads' longest MEMs occurred only in the sequences from which those reads were generated (``true positive'' reads).  For some parameter settings we achieved significant compression while only slightly decreasing the true-positive rate.
\end{abstract}

\newpage

\section{Introduction}
\label{sec:introduction}

\begin{figure}[t]
\begin{center}
\includegraphics[width=.4\textwidth]{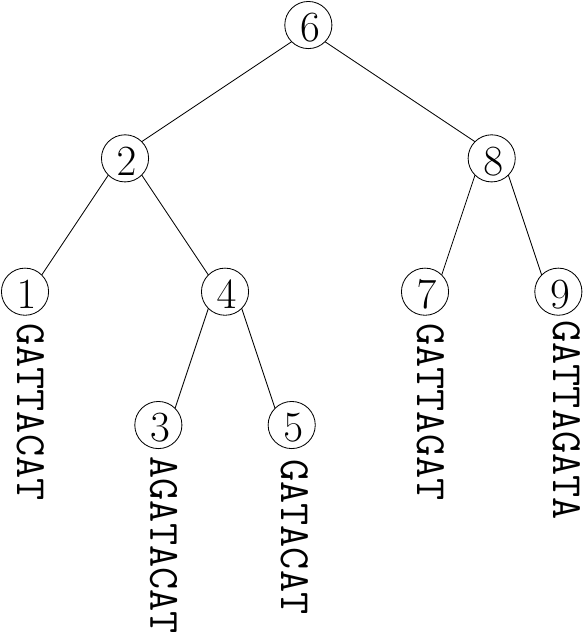}

\bigskip

\resizebox{.35\textwidth}{!}
{\begin{tabular}{ccc}
{\tt ACA}: 2 & {\tt ATT}: 6 & {\tt TAC}: 2 \\
{\tt AGA}: 6 & {\tt CAT}: 2 & {\tt TAG}: 8 \\
{\tt ATA}: 6 & {\tt GAT}: 6 & {\tt TTA}: 6
\end{tabular}}

\caption{A toy phylogenetic tree {\bf (top)} with Kraken's $k$-mer index for $k = 3$ {\bf (bottom)}.}
\label{fig:kraken}
\end{center}
\end{figure}

Kraken~\cite{wood2014kraken} is probably the best-known metagenomic tool for taxonomic classification.  Given a phylogenetic tree for a collection of genomes and a value $k$, it stores an index mapping each $k$-mer in the collection to the root of the lowest subtree containing all occurrences of that $k$-mer.  Later, given a DNA read --- which may not match exactly in any of genomes in the collection --- it tries to map all the $k$-mers in that read to subtrees in the tree and then to choose a small subtree likely to contain the source of the read.  For example, if Kraken is given the toy phylogenetic tree shown at the top of Figure~\ref{fig:kraken} and $k = 3$, then it will store the $k$-mer index shown at the bottom of that figure.  Later, given the toy read {\tt ATAC}, it will map {\tt ATA} to 6 and {\tt TAC} to 2.  Since the subtree rooted at 6 contains the one rooted at 2, it will report that the read probably came from a genome in the subtree rooted at 2.

Nasko et al.~\cite{nasko2018refseq} showed that a static choice of $k$ is problematic, since “the [reference] database composition strongly influence[s] the performance”, with larger $k$ values  working better as the collection of genomes grows over time.  Limiting all analyses to a single choice of $k$ causes other problems as well.  First, some branches of the taxonomic tree are well studied and contain a large number of genome assemblies for diverse strains and species.  Other branches are scientifically significant but harder to study, and contain only a few genomes.  In the more richly sampled spaces, larger values of $k$ will better allow for discrimination at deeper levels of the tree.

Choosing a constant value for $k$ also conflicts with the varying error rates across sequencing technologies.  For the high-accuracy Illumina technology, we expect longer matches to the data base and should favour a larger $k$.  For a high-error-rate technology like Oxford Nanopore, we expect shorter matches and a small $k$ is better.  To this end, many widely tools for classifying long (error-prone) reads use matching statistics and/or full-text indexes~\cite{kovaka2021targeted,ahmed2021pan}, as do some for short reads~\cite{kim2016centrifuge,menzel2016fast}.  Nasko et al.\ observed that
\begin{quotation}
``alternative approaches to traditional $k$-mer-based [lowest common ancestor] identification methods, such as those featured within KrakenHLL~\cite{breitwieser2018krakenuniq}, Kallisto~\cite{bray2016near}, and DUDes~\cite{piro2016dudes}, will be required to maximize the benefit of longer reads coupled with ever-increasing reference sequence databases and improve sequence classification accuracy.''
\end{quotation}

Cheng et al.~\cite{cheng??factors} showed that finding the maximal exact matches (MEMs) of the read with respect to the collection and then mapping each MEM to the root of the lowest subtree containing all occurrences of that MEM, gives better results than mapping $k$-mers for any single $k$.  However, they did not give a space- and time-efficient index for finding and mapping MEMs.  As a potential step toward working with MEMs, Gagie et al.~\cite{gagie2022katka} described an LZ77-based index KATKA that takes $O (z \log n)$ space, where $z$ is the number of phrases in the LZ77 parse of the collection of genomes and $n$ is the total length of the collection, and works like Kraken but taking $k$ at query time instead of at construction time.

KATKA finds the indices of the genomes containing the first and last occurrences of each $k$-mer in the collection, then performs a lowest common ancestor (LCA) query on those genomes in the tree to find the root of the smallest subtree containing all the occurrences of that $k$-mer.  As far as we know, however, there is no practical way to find MEMs with LZ77- or grammar-based indexes, even if there have been some promising recent developments~\cite{gao2022computing,navarro2023computing} in this direction.  Thus, KATKA is not yet a practical implementation of Cheng et al.'s idea.

Since an LCA data structure for the phylogenetic tree takes a constant number of bits per genome, the main challenge to implementing Cheng et al.'s idea is to find the MEMs of the read with respect to the collection and then to find the genomes containing the first and last occurrence of each MEM.  We call all this information the {\em MEM table} for the read.  We describe in Section~\ref{sec:preliminaries} how we can extend a technique by Ohlebusch et al.~\cite{ohlebusch2010computing} to build the MEM table in constant time per character in the read plus $O (\log n)$ time per MEM as long as we are willing to use an $O (n)$-bit augmented FM-index for the collection --- but a space usage of $O (n)$ bits is prohibitive when the collection is large and anyway wasteful when it is highly repetitive.  The most practical way we know of to build the MEM table is with C\'aceres and Navarro's~\cite{caceres2022faster} block-tree compressed suffix tree, but that offers more functionality than we need at the cost of using more space than we would like (``1--3 bits per symbol in highly repetitive text collections'').

In this paper we build approximations of MEM tables using augmented FM-indexes over
\begin{itemize}
\item a string kernel for the collection,
\item a minimizer digest for the collection,
\item a string kernel for a minimizer digest for the collection.
\end{itemize}
String kernels and minimizer digests are lossily compressed representations of strings, which we review in Section~\ref{sec:preliminaries}.  We need a special kind of string kernel that we call a KATKA kernel and define also in Section~\ref{sec:preliminaries}.  We can use KATKA kernels and minimizer digests to reduce the size of the augmented FM-index, at the cost of limiting the lengths of matches and reporting some false-positive matches.  To test how we can trade off accuracy for compression, we built augmented FM-indexes over a test dataset and KATKA kernels, minimizer digests, and KATKA kernels of minimizer digests for that dataset with various parameter settings, and checked for how many of a set of simulated reads their longest MEMs occurred only in the sequences from which those reads were generated (``true positive'' reads).  For some parameter settings we achieved significant compression while only slightly decreasing the true-positive rate.

\section{Preliminaries}
\label{sec:preliminaries}

\subsection{Augmented FM-indexes}

Ohlebusch et al.~\cite{ohlebusch2010computing} showed how, if we store an augmented FM-index, then when given a read we can find its MEMs quickly.  We first show how to extend their technique to computing the MEM table in constant time per character in the read and $O (\log n)$ time per MEM.

Suppose each genome in the collection is terminated by a special separator character {\tt \$} as shown in Figure~\ref{fig:FM}.  The augmented FM-index consists of data structures supporting access, rank and select on the collection's Burrows-Wheeler Transform (BWT)\footnote{To reduce the size of the figure we have actually shown the genomes' extended BWT~\cite{mantaci2007extension}, which is functionally equivalent as far as we are concerned as long as each genomes has length $\Omega ( \log n)$.  Notice some LCP values, such as $\LCP [4]$, ``wrap around'' and count a character in the BWT.}; access, range-minimum and range-maximum on their suffix array (SA); range-minimum, range-maximum, previous smaller value (PSV) and next smaller value (NSV) queries on their longest common prefix (LCP) array; and rank on the bitvector $B$ with a 1 marking each {\tt \$} in the collection.  As long as the collection is over a constant-size alphabet, these data structures together take $O (n)$ bits with all their queries taking at most $O (\log n)$ time.  They are also implemented in the Succinct Data Structure Library (SDSL)~\cite{gog2014theory} as components of a compressed suffix tree.

\begin{figure}[t]
\begin{center}
\resizebox{\textwidth}{!}
{\begin{tabular}{c@{\hspace{5ex}}c}
\begin{tabular}{r|rccl}
$i$ & $\SA [i]$ & $\LCP [i]$ & $\BWT [i]$ & context\\
\hline
 0 & 17 & 0 & \tt T  & \tt $\mathtt{\$}$AGATACA\\
 1 & 25 & 1 & \tt T  & \tt $\mathtt{\$}$GATACA\\
 2 & 8 & 4 & \tt T  & \tt $\mathtt{\$}$GATTACA\\
 3 & 34 & 6 & \tt T  & \tt $\mathtt{\$}$GATTAGA\\
 4 & 44 & 9 & \tt A  & \tt $\mathtt{\$}$GATTAGAT\\
 5 & 43 & 0 & \tt T  & \tt A$\mathtt{\$}$GATTAGA\\
 6 & 13 & 1 & \tt T  & \tt ACAT$\mathtt{\$}$AGA\\
 7 & 21 & 5 & \tt T  & \tt ACAT$\mathtt{\$}$GA\\
 8 & 4 & 8 & \tt T  & \tt ACAT$\mathtt{\$}$GAT\\
 9 & 30 & 1 & \tt T  & \tt AGAT$\mathtt{\$}$GAT\\
10 & 39 & 4 & \tt T  & \tt AGATA$\mathtt{\$}$GAT\\
11 &  9 & 5 & \tt $\mathtt{\$}$ & \tt AGATACAT\\
12 & 15 & 1 & \tt C  & \tt AT$\mathtt{\$}$AGATA\\
13 & 23 & 3 & \tt C  & \tt AT$\mathtt{\$}$GATA\\
14 &  6 & 6 & \tt C  & \tt AT$\mathtt{\$}$GATTA\\
15 & 32 & 8 & \tt G  & \tt AT$\mathtt{\$}$GATTA\\
16 & 41 & 2 & \tt G  & \tt ATA$\mathtt{\$}$GATTA\\
17 & 11 & 3 & \tt G  & \tt ATACAT$\mathtt{\$}$A\\
18 & 19 & 7 & \tt G  & \tt ATACAT$\mathtt{\$}$\\
19 &  1 & 2 & \tt G  & \tt ATTACAT$\mathtt{\$}$\\
20 & 27 & 4 & \tt G  & \tt ATTAGAT$\mathtt{\$}$\\
21 & 36 & 7 & \tt G  & \tt ATTAGATA$\mathtt{\$}$\\
22 & 14 & 0 & \tt A  & \tt CAT$\mathtt{\$}$AGAT
\end{tabular}
&
\begin{tabular}{r|rccl}
$i$ & $\SA [i]$ & $\LCP [i]$ & $\BWT [i]$ & context\\
\hline
23 & 22 & 4 & \tt A  & \tt CAT$\mathtt{\$}$GAT\\
24 &  5 & 7 & \tt A  & \tt CAT$\mathtt{\$}$GATT\\
25 & 31 & 0 & \tt A  & \tt GAT$\mathtt{\$}$GATT\\
26 & 40 & 3 & \tt A  & \tt GATA$\mathtt{\$}$GATT\\
27 & 10 & 4 & \tt A  & \tt GATACAT$\mathtt{\$}$\\
28 & 18 & 8 & \tt $\mathtt{\$}$ & \tt GATACAT\\
29 &  0 & 3 & \tt $\mathtt{\$}$ & \tt GATTACAT\\
30 & 26 & 5 & \tt $\mathtt{\$}$ & \tt GATTAGAT\\
31 & 35 & 8 & \tt $\mathtt{\$}$ & \tt GATTAGATA\\
32 & 16 & 0 & \tt A  & \tt T$\mathtt{\$}$AGATAC\\
33 & 24 & 2 & \tt A  & \tt T$\mathtt{\$}$GATAC\\
34 &  7 & 5 & \tt A  & \tt T$\mathtt{\$}$GATTAC\\
35 & 33 & 7 & \tt A  & \tt T$\mathtt{\$}$GATTAG\\
36 & 42 & 1 & \tt A  & \tt TA$\mathtt{\$}$GATTAG\\
37 & 12 & 2 & \tt A  & \tt TACAT$\mathtt{\$}$AG\\
38 & 20 & 6 & \tt A  & \tt TACAT$\mathtt{\$}$G\\
39 &  3 & 8 & \tt T  & \tt TACAT$\mathtt{\$}$GA\\
40 & 29 & 2 & \tt T  & \tt TAGAT$\mathtt{\$}$GA\\
41 & 38 & 5 & \tt T  & \tt TAGATA$\mathtt{\$}$GA\\
42 &  2 & 1 & \tt A  & \tt TTACAT$\mathtt{\$}$G\\
43 & 28 & 3 & \tt A  & \tt TTAGAT$\mathtt{\$}$G\\
44 & 37 & 6 & \tt A  & \tt TTAGATA$\mathtt{\$}$G\\
&&&&
\end{tabular}
\end{tabular}}

\vspace{5ex}

\setlength{\tabcolsep}{2pt}
\begin{tabular}{c}
\begin{tabular}{ccccccccccccccccccccccccccccccccccccccccccccc}
\tt G & \tt A & \tt T & \tt T & \tt A & \tt C & \tt A & \tt T & \tt \$ & \hspace{2ex} &
\tt A & \tt G & \tt A & \tt T & \tt A & \tt C & \tt A & \tt T & \tt \$ & \hspace{2ex} &
\tt G & \tt A & \tt T & \tt A & \tt C & \tt A & \tt T & \tt \$ \\
0 & 1 & 2 & 3 & 4 & 5 & 6 & 7 & 8 &  & 9 & 10 & 11 & 12 & 13 & 14 & 15 & 16 & 17 & & 18 & 19 & 20 & 21 & 22 & 23 & 24 & 25
\end{tabular} \\[3ex]
\begin{tabular}{ccccccccccccccccccccccccccccccccccccccccccccc}
\tt G & \tt A & \tt T & \tt T & \tt A & \tt G & \tt A & \tt T & \tt \$ & \hspace{2ex} &
\tt G & \tt A & \tt T & \tt T & \tt A & \tt G & \tt A & \tt T & \tt A & \tt \$ \\
26 & 27 & 28 & 29 & 30 & 31 & 32 & 33 & 34 & & 35 & 36 & 37 & 38 & 39 & 40 & 41 & 42 & 43 & 44
\end{tabular}
\end{tabular}
\setlength{\tabcolsep}{6pt}

\vspace{5ex}

$B = \mathrm{000000001000000001000000010000000010000000001}$
\end{center}
\caption{The augmented FM-index for our toy collection of genomes.}
\label{fig:FM}
\end{figure}

Given the read {\tt ACATA}, for example, we start a backward search with BWT interval $\BWT [0..44]$ (the entire BWT).  After 3 backward steps we find the interval $\BWT [16..18]$ for {\tt ATA}.  Since this interval does not contain a copy of the preceding character {\tt C} in the read, we know {\tt ATA} is a MEM of {\tt ACATA} with respect to the collection.  We use range-minimum and range-maximum queries over $\SA [16..18]$ and access to SA to determine that the first and last occurrences of {\tt ATA} start at positions 11 and 41 in the collection.  Since $B.\rank_1 (11) = 1$ and $B.\rank_1 (41) = 4$, we know those occurrences are in the second and fifth genomes in the collection (stored at nodes 3 and 9 in the phylogenetic tree).  Notice we consider only the first and last occurrences and not the occurrence starting at position 19, for example.

We then use rank and select queries on the BWT to look for the previous copy $\BWT [14]$ of {\tt C} and next copy of {\tt C} (which does not exist); use a range-minimum query on $\LCP [14 + 1 = 15..16]$ to find the position 16 of the length 2 of the longest prefix {\tt AT} of {\tt ATA} that is preceded by {\tt C} in the collection; use access to the LCP to retrieve that value 2; and use $\PSV (16) = 12$ and $\NSV (16) = 22$ queries to find the interval $\BWT [12..22 - 1 = 21]$ for that prefix {\tt AT}.  After 2 backward steps we find the interval $\BWT [6..8]$ for {\tt ACAT}.  We use range-minimum and range-maximum queries over $\SA [6..8]$ and access to SA to determine that the first and last occurrences of {\tt ACAT} start at positions 4 and 21 in the collection.  Since $B.\rank_1 (4) = 0$ and $B.\rank_1 (21) = 2$, we know those occurrences are in the first and third genomes in the collection (stored at nodes 1 and 5 in the phylogenetic tree).

\subsection{String kernels and KATKA kernels}

Ferrada, Gagie, Hirvola and Puglisi~\cite{ferrada2014hybrid,gagie2015searching} and Prochazka and Holub~\cite{prochazka2014compressing} (see also~\cite{ferrada2018hybrid}) independently defined the order-$k_{\max}$ kernel of a string to be the subsequence consisting of the characters in the first occurrence of any distinct $k_{\max}$-mer in the string, with maximal omitted substrings replaced by copies of a new separator character~{\tt \#}.  Since we want to find the first and last occurrences of matches, we define the {\em order-$k_{\max}$ KATKA kernel} of a collection of genomes essentially the same way, but with the subsequence consisting of the characters in the first {\em or last} occurrence of any distinct $k_{\max}$-mer in the string, and the copies of the separator character {\tt \$}.  Since reads will not contain {\tt \$}, we also do not replace with~{\tt \#} all maximal omitted substrings adjacent to copies of~{\tt \$}.

By construction, for $k \leq k_{\max}$, every $k$-mer from the normal alphabet (so not including {\tt \$}) in the original string occurs in the KATKA kernel and vice versa.  Moreover, if there are $i$ copies of {\tt \$} to the left of the first occurrence of such a $k$-mer in the kernel, then the first occurrence of that $k$-mer in the collection is in the $(i + 1)$st genome (and symmetrically for the last occurrences).  The running example we have used so far is too small to illustrate properly the advantages and disadvantages of KATKA kernels, so Figure~\ref{fig:genomes} shows a slightly larger collection of slightly longer toy genomes and Figure~\ref{fig:4-mers} shows the subsequence consisting of the characters in the first or last occurrence of each distinct 4-mer and the copies of {\tt \$}.  Figure~\ref{fig:kernel} shows the 4th-order KATKA kernel of the collection with maximal omitted substrings replaced by copies of {\tt \#}.  In this example, the 4th-order KATKA kernel is about half the size of the original collection, but this varies in practice depending on $k_{\max}$ and the size and repetitiveness of the collection.  The 5th-order KATKA kernel, which we do not show, is about 70\% of the size of the original collection.

\begin{figure}[p]
\resizebox{\textwidth}{!}
{\tt \begin{tabular}{l}
ACTTAGCTGACGTTCCGGGTGTTTTTGGCCATCTTCTATAGATTTCCCAGAGACATACTAGGCGTGCTGAAGTTGTGACTCGCGGCCGTATTTTCTAACG\$ \\
ACTTAGCTGACGTTCCGGGTGTTTTAGGCCATCTTCTATAGATTTCTCAGAGACATAGTAGGCGTGCTGAAGTTGTGACTCGCGGCCGTATTCCCTAACG\$ \\
ACTTAGCTGACGTTCCGGGTGTTTTAGGCCATCTTCTATAGTTTTCTCAGAGACATACTAGGCGTGCTGAAGTTGTCACGCGCGCCCGTATTTCCTAACG\$ \\
ACTTAGCTGACGTTCAGGGTGTTTTAGGCCATCTTCTATAGTTTTCTCAGAGACATAGTAGGCGTGCTGAAGTTGTCACTCGCGCCCGTATTTCCTAACG\$ \\
TCAGAGCTGAGGTTCGGGGTGATTTAGGACATCTTCCATCGATTTCTCAGAGACGTCCTCAGCGTGCTCAAGTTGTCACCCGCGGCCGTTATTCCTAACG\$ \\
TCCGAGCTGAGGTTCGGGGTGGTTTAGGTCATCTTCTATAAATTTCTCAGAGACGTCCCCAGCGTGCTCAAGTTGTCACTCGCGGCCGTTTTTCCGAACG\$ \\
TCATAGCTGAGGTACGGGGTGGTTTAGGCCAGCTTCTATAGATTTCTCAGACACGAGCAGGGCGTGCTTAAGTTGTCACTCGCGGCCGTTTTTCCTAACG\$ \\
TCATAGCTGAGGTACGGGGTGGTTTAGGCCACCTTCTATAGATTTGTCAGACACGAGCTCGGCGTGCTGAAGTTGTCACCCGCGGCCGTTTTTCCTAACG\$ \\
TCCAAGCGTCCGTTCGGGGTGGGTTAGGCGATCTTCTGTAGAGTTCTCGGAGACAAGCTAGGCGTGCTGATGTTGTCATTCGCGGCCGTGTTCCCTAACG\$ \\
TCCAAGCTTCCGTTCGGGGTGGGTTAGACGATCTTCTGTACAGTTCTTTGAGACAAGCTAGGCGTGCTGAAGTTGTCACACGCGGCCGTGTTCCCTAACG\$ \\
TGACAGCGGACGTTCGGGGTGGGTTAGGACATCTTCCGTAGATTTCTCGGATACAAGCTAGGCGTTCTGAAGTTGGCACTCGCGGCCTTGTTCCCTAACG\$ \\
TCCCTGCTGACGATCGGGGTAGGTTAGGACATCTTCCGTTGATTTCTCGGATACAAGCTCGGCGTTCTGAAGTTGGCACTCGCGGCCGTGTTCCCTAACG\$ \\
TAATATCAGACGTTCGGGCTGGGCTAGTCCATCTTCTTTAGATTTCTCAGAGACTTGCTAGGCGTGCTGAAGTTGGCACTCGTGGCCGTGTTCCCTAACG\$ \\
TAATATCAGACGTTCGGGATGGGCTAGTCCATCTTCTTTAGATTTCTCAGAGACATGCTAGGCGTGCTGCAGTTGTCACTCGTGGCCGTGTTCCGTTACG\$ \\
TAATATCAGACGTTCGGGCTGGATTAGGCCATCTTCTTTAGATTTCTCAGAGACATGCTAGGCGTGCTGAAGTTGGTAATCGCGGCCCTGTTCTTTAACG\$ \\
TAATATCAGACGTTCGGGCCGGGTTAGGCCATCTTCTTTAGATTTCTCAGAGACATGCTAGGCGTGCTGAAGTTGGCAATCGCGGACCTGTTCTCTAACG\$
\end{tabular}}
\caption{A slightly larger collection of slightly longer toy genomes.}
\label{fig:genomes}
\end{figure}

\begin{figure}[p]
\resizebox{\textwidth}{!}
{\tt \begin{tabular}{l}
ACTTAGCTGACGTTCCGGGTGTTTTTGGCCATCTTCTATAGATTTCCCAGAGACATACTAGGCGTGCTGAAGTTGTGACTCGCGGCCGTATT\ \ CTAACG\$ \\
\ \ \ \ \ \ \ \ \ \ \ \ \ \ \ \ \ \ \ \ \ \ TTTA\ \ \ \ \ \ \ \ \ \ \ \ \ \ \ \ \ \ TCTCAG\ \ \ \ \ TAGTAG\ \ \ \ \ \ \ \ \ \ \ \ TGTG\ \ \ \ \ \ \ \ \ \ \ \ ATTCCCTA\ \ \ \$ \\
\ \ \ \ \ \ \ \ \ \ \ \ \ \ \ \ \ \ \ \ \ \ \ \ \ \ \ \ \ \ \ \ \ \ \ \ \ \ \ \ \ \ \ \ \ \ \ \ \ \ \ \ \ \ \ TACTA\ \ \ \ \ \ \ \ \ \ \ \ \ TGTCACGCGCGCCCG\ \ \ \ TCCT\ \ \ \ \$ \\
\ \ \ \ \ \ \ \ \ \ \ \ TTCAGGG\ \ \ \ \ \ \ \ \ \ \ \ \ \ \ \ \ \ \ \ \ \ \ \ \ \ \ \ \ \ \ \ \ \ \ \ \ AGTA\ \ \ \ \ \ \ \ \ \ \ \ \ \ \ \ CACT\ GCGCC\ GTAT\ \ \ \ \ \ \ \ \ \$ \\
\ \ \ GAGCTGAGGTTCGGGGTGAT\ \ AGGAC\ \ \ \ TCCATCGAT\ \ \ \ \ \ \ \ \ \ CGTCCTCAGCG\ GCTCAAG\ \ \ \ CACCCGC\ \ \ \ GTTATT\ \ \ \ \ \ \ \$ \\
\ CCGAG\ \ \ \ \ \ \ \ \ \ \ \ GTGGT\ \ \ GGTCAT\ \ \ \ \ ATAAATT\ \ \ \ \ \ \ \ \ \ \ \ CCCCA\ \ \ \ \ \ TCAA\ \ \ \ \ \ \ \ \ \ \ \ \ \ \ \ \ \ \ \ \ \ CCGAAC\ \$ \\
\ \ \ \ \ \ \ \ \ \ GGTACGG\ \ \ \ \ \ \ \ \ \ \ CCAGCTT\ \ \ \ \ \ \ \ \ \ \ \ \ \ \ ACACGAGCAGGGC\ \ \ CTTAAG\ \ \ \ \ \ \ \ \ \ \ \ \ \ \ \ \ \ \ \ \ \ \ \ \ \ \ \ \$ \\
\ CATAGC\ GAGG\ ACGG\ \ \ \ \ \ \ \ \ \ \ CCACCTTCTATAG\ \ \ \ \ \ \ \ \ \ \ \ CGAGC\ \ \ \ \ \ \ \ \ \ \ \ \ \ \ \ \ \ CACCCGC\ \ \ \ GTTTTTCCT\ \ \ \ \$ \\
\ CCAAGCGTC\ \ \ \ \ \ \ \ \ TGGG\ \ \ \ GCGATC\ TCTGTAGAGT\ \ \ CGGAGACAAGCTA\ \ \ \ \ \ \ \ GATGT\ \ TCATTC\ \ \ \ \ \ \ \ \ \ \ \ \ \ \ \ \ \ \ \$ \\
\ CCAAGCTT\ \ \ \ \ \ \ \ \ \ \ \ \ \ \ \ \ \ \ \ \ \ \ \ \ \ \ TGTACAGT\ CTTTGAG\ \ \ \ \ \ \ \ \ \ \ \ \ \ \ \ \ \ \ \ \ \ \ \ CACACGC\ \ \ \ \ \ \ \ \ \ \ \ \ \ \ \ \ \$ \\
\ \ ACAGCG\ \ \ \ \ \ \ \ \ GGTG\ \ \ \ \ \ \ \ \ \ \ \ \ \ \ CGTA\ \ \ \ \ \ \ \ GGATA\ \ \ \ \ \ \ \ \ \ \ \ \ \ \ \ \ \ \ \ \ GGCAC\ \ \ \ \ GCCTTG\ \ \ \ \ \ \ \ \ \ \$ \\
\ \ CCTGCTGACGATCGGGGTAGGT\ AGGA\ \ \ \ \ \ \ \ \ TTGAT\ \ \ \ \ \ GATACAAGCTC\ \ \ \ \ TCTG\ \ \ \ \ \ \ \ \ \ \ \ \ \ \ \ \ \ \ \ \ \ \ \ \ \ \ \ \ \ \ \$ \\
TAATATCA\ \ \ \ \ \ \ \ GGCTGG\ \ \ AGTC\ \ \ \ \ \ \ \ \ \ \ \ \ \ \ \ \ \ \ \ \ \ GACTTGC\ \ \ \ \ \ \ \ \ \ \ \ \ \ \ \ \ GCACTCGT\ \ \ \ \ \ \ \ TCCCTA\ \ \ \$ \\
\ \ \ \ \ \ \ \ \ \ \ \ \ \ \ GGGATGGG\ TAGTCCA\ \ \ \ \ \ \ \ \ \ \ \ \ \ \ \ \ \ \ \ \ \ CATGC\ \ \ \ \ \ \ \ CTGCAGTTGTCACTCGTGG\ \ GTGTTCCGTTACG\$ \\
\ \ \ \ \ \ \ \ \ \ \ \ \ \ \ \ GGCTGGATTA\ \ \ \ \ \ \ \ \ \ \ \ \ \ \ \ \ \ \ \ \ \ \ \ \ \ \ \ \ \ \ \ \ \ \ \ \ \ \ \ \ \ \ \ \ \ \ TGGTAATC\ CGGCCCT\ \ \ \ \ TTAA\ \ \$ \\
TAATATCAGACGTTCGGGCCGGGTTAGGCCATCTTCTTTAGATTTCTCAGAGACATGCTAGGCGTGCTGAAGTTGGCAATCGCGGACCTGTTCTCTAACG\$
\end{tabular}}
\caption{The subsequence consisting of the characters in the first or last occurrence of each distinct 4-mer and the copies of {\tt \$}, with omitted characters replaced by spaces.}
\label{fig:4-mers}
\end{figure}

\begin{figure}[p]
\resizebox{\textwidth}{!}
{\tt \begin{tabular}{l}
ACTTAGCTGACGTTCCGGGTGTTTTTGGCCATCTTCTATAGATTTCCCAGAGACATACTAGGCGTGCTGAAGTTGTGACTCGCGGCCGTATT\#CTAACG\$T \\
TTA\#TCTCAG\#TAGTAG\#TGTG\#ATTCCCTA\$TACTA\#TGTCACGCGCGCCCG\#TCCT\$TTCAGGG\#AGTA\#CACT\#GCGCC\#GTAT\$GAGCTGAGGTTCG \\
GGGTGAT\#AGGAC\#TCCATCGAT\#CGTCCTCAGCG\#GCTCAAG\#CACCCGC\#GTTATT\$CCGAG\#GTGGT\#GGTCAT\#ATAAATT\#CCCCA\#TCAA\#CCGA \\
AC\$GGTACGG\#CCAGCTT\#ACACGAGCAGGGC\#CTTAAG\$CATAGC\#GAGG\#ACGG\#CCACCTTCTATAG\#CGAGC\#CACCCGC\#GTTTTTCCT\$CCAAGC \\
GTC\#TGGG\#GCGATC\#TCTGTAGAGT\#CGGAGACAAGCTA\#GATGT\#TCATTC\$CCAAGCTT\#TGTACAGT\#CTTTGAG\#CACACGC\$ACAGCG\#GGTG\#C \\
GTA\#GGATA\#GGCAC\#GCCTTG\$CCTGCTGACGATCGGGGTAGGT\#AGGA\#TTGAT\#GATACAAGCTC\#TCTG\$TAATATCA\#GGCTGG\#AGTC\#GACTTG \\
C\#GCACTCGT\#TCCCTA\$GGGATGGG\#TAGTCCA\#CATGC\#CTGCAGTTGTCACTCGTGG\#GTGTTCCGTTACG\$GGCTGGATTA\#TGGTAATC\#CGGCCC \\
T\#TTAA\$TAATATCAGACGTTCGGGCCGGGTTAGGCCATCTTCTTTAGATTTCTCAGAGACATGCTAGGCGTGCTGAAGTTGGCAATCGCGGACCTGTTCT \\
CTAACG\$
\end{tabular}}
\caption{The subsequence consisting of the characters in the first or last occurrence of each distinct 4-mer --- the 4th-order KATKA kernel --- and the copies of {\tt \$}, with maximal omitted substrings replaced by copies of {\tt \#}, except for those adjacent to~{\tt \$}.}
\label{fig:kernel}
\end{figure}

\subsection{Minimizer digests}

To build a {\em minimizer digest}~\cite{roberts2004reducing} for a string $S [1..n]$, we
\begin{enumerate}
\item choose parameters $k$ and $w$ and a hash function $h (\cdot)$ function on $k$-mers,
\item mark each $k$-mer $S [j..j + k - 1]$ in $S$ such that $h (S [j..j + k - 1])$ is the leftmost occurrence of the minimum in $h (S [i..i + k - 1], \ldots, S [i + w - 1..(i + w - 1) + k - 1])$ for some $i$ with $i \leq j < i + w$,
\item return the sequence of marked $k$-mers' hashes.
\end{enumerate}
For example, suppose $k = 3$, $w = 10$ and the hash function $h (\cdot)$ takes a triple over $\{\mathtt{A}, \mathtt{C}, \mathtt{G}, \mathtt{T}\}$ as a 3-digit number $x$ in base 4 and returns $(2544 x + 3937) \bmod 8863$.  The minimizer digests for the toy genomes in Figure~\ref{fig:genomes} (excluding {\tt \$}s) are shown in Figure~\ref{fig:digests} separated by {\tt \$}s and with the 64 triples over $\{\mathtt{A}, \mathtt{C}, \mathtt{G}, \mathtt{T}\}$ mapped to ASCII values between 37 and 100.  Minimizer digests are widely used in bioinformatics to reduce tools' time and space requirements; for example, they are used this way in Kraken 2~\cite{wood2019improved}, mdBG~\cite{ekim2021minimizer} and SPUMONI 2~\cite{ahmed2023spumoni}.

We note that although the first minimizer digest \verb$=c<J_cA\2X<G2@'cKNJX5$ is 21 characters while the first genome is 100 characters, the digest is over an alphabet of size 64 instead of 4; therefore, the minimizer is 126 bits while the genome is 200 bits.  The space of the auxiliary data structures for an augmented FM-index for the minimizer digest still depends on the number 21 of characters in the digest, however.

\begin{figure}[t]
\begin{center}
\begin{BVerbatim}
=c<J_cA\2X<G2@'cKNJX5$=c<J_cA\2X\G3K@'cKNJ<5$=c<J_cA\2X\G2@'C6J<
5$=c_cA\2X\G3K@'C6J<5$G__/\<.GC<@CJJ<5$.__CXUGC<@CNJ<.'$G___c=\2
X\.+@CNJ<5$G___92XC.G@'CJJ<5$<<J__.\GG2@QCNJ<5$<<J__.\\\G2@'CNJ<
5$NN__\<J\N2\'+N<5$<.__\<J\N/=N'+J<5$XcN2C<\\\G@2@'+J<5$XcNQC<\\
\GQ2@+C6J<$XcN/A\\\GQ2@'_N\5$XcNJ_\\\GQ2@'+925$
\end{BVerbatim}
\caption{Minimizer digests for the toy genomes in Figure~\ref{fig:genomes}, separated by {\tt \$}s.}
\label{fig:digests}
\end{center}
\end{figure}

We say the concatenation of the minimizer digests for the genomes in a collection, separated by {\tt \$}s, is the minimizer digest for the collection.  By construction, if $\alpha$ is the minimizer digest for a pattern and there are $i$ copies of {\tt \$} to the left of the first occurrence of $\alpha$ in the minimizer digest for the collection, then the first occurrence of the pattern cannot be before the $(i + 1)$st genome (and symmetrically for the last occurrences) --- although the pattern may not occur in that genome and possibly not in the whole collection.

\subsection{KATKA kernels of minimizer digests}
\label{subsec:kernels_of_digests}

Of course, we can also build KATKA kernels of minimizer digests.  Figure~\ref{fig:minimizer_kernel} shows the subsequence consisting of the characters in the first or last occurrence of each distinct pair --- the 2nd-order KATKA kernel --- and the copies of {\tt \$} in Figure~\ref{fig:digests}, with maximal omitted substrings replaced by copies of {\tt \#}.  It consists of 220 6-bit characters (1320 bits) plus the 16 {\tt \$}s; the original minimizer digest consists of 287 6-bit characters (1722 bits) plus the {\tt \$}s, the 4th-order KATKA kernel consists of 798 2-bit characters (1596 bits) plus the {\tt \$}s, and the collection of toy genomes itself consists of 1600 2-bit characters (3200 bits) plus the {\tt \$}s.  We note that pairs of minimizers with $k = 3$ and $w = 10$ can represent substrings as short as 4 characters or as long as 17 characters in the genomes; in our example, on average a pair of minimizers represents about $2 \cdot (1600 / 287) \approx 11.15$ characters.

\begin{figure}[t]
\begin{center}
\begin{BVerbatim}
=c<J_cA\2X<G2@'cKNJX5$X\G3K@'cKNJ<5$c<#'C6J$=c_cA#G3K@$G__/\<.GC
<@CJJ$.__CXUGC<@CN#.'$_c=\2X\.+@C$G_#_92XC.G@#CJJ$<<#_.\GG#@QC$<
<#_.\\#G2#'CN$NN__\#J\N2\'+N<$<.__\<J\N/=N'+J$XcN2C<\#G@2#+J<5$N
QC<\#GQ2@+C6J<$N/A\#'_N\5$XcNJ_\\\GQ2@'+925$
\end{BVerbatim}
\caption{The subsequence consisting of the characters in the first or last occurrence of each distinct pair --- the 2nd-order KATKA kernel --- and the copies of {\tt \$} in Figure~\ref{fig:digests}, with maximal omitted substrings replaced by copies of {\tt \#}.}
\label{fig:minimizer_kernel}
\end{center}
\end{figure}

KATKA kernels of minimizer digests may inherit the strengths of both: with kernelization we can take advantage of repetition to compress, while using minimizers allows us to keep the parameter $k$ in the kernelization small while still dealing with reasonably long patterns.

\section{Approximating MEM tables with FM-indexes of KATKA kernels and minimizer digests}
\label{sec:approximating}

Once we have built a KATKA kernel or minimizer digest for a collection of genomes, or a KATKA kernel of a minimizer digest, we can build an augmented FM-index over it.  For example, Figure~\ref{fig:indexes} shows the first and last lines of the augmented FM-indexes for the 4th-order KATKA kernel in Figure~\ref{fig:kernel}; the minimizer digest in Figure~\ref{fig:digests}; and the 2nd-order KATKA kernel of the minimizer digest, from Figure~\ref{fig:minimizer_kernel}.  In all three cases, we include an implicit end-of-file character less than any other.

\begin{figure}[p]
\begin{center}
\resizebox{\textwidth}{!}
{\begin{tabular}{c@{\hspace{5ex}}c}
\begin{tabular}{r|rccl}
$i$ & $\SA [i]$ & $\LCP [i]$ & $\BWT [i]$ & context\\
\hline
0 & 815 & 0 & \tt \$ & \\
1 & 321 & 0 & \tt T & \tt \#ACACGAGCA\dots \\
2 & 354 & 3 & \tt G & \tt \#ACGG\#CCAC\dots \\
3 & 550 & 2 & \tt T & \tt \#AGGA\#TTGA\dots \\
4 & 209 & 5 & \tt T & \tt \#AGGAC\#TCC\dots \\
5 & 167 & 3 & \tt G & \tt \#AGTA\#CACT\dots \\
\vdots & \vdots & \vdots & \vdots & \vdots
\end{tabular}
&
\begin{tabular}{r|rccl}
$i$ & $\SA [i]$ & $\LCP [i]$ & $\BWT [i]$ & context\\
\hline
\vdots & \vdots & \vdots & \vdots & \vdots \\
810 & 477 & 3 & \tt C & \tt TTTGAG\#CAC\dots \\
811 & 23 & 4 & \tt T & \tt TTTGGCCATC\dots \\
812 & 390 & 3 & \tt T & \tt TTTTCCT\$CC\dots \\
813 & 22 & 4 & \tt T & \tt TTTTGGCCAT\dots \\
814 & 389 & 4 & \tt G & \tt TTTTTCCT\$C\dots \\
815 & 21 & 5 & \tt G & \tt TTTTTGGCCA\dots
\end{tabular}
\end{tabular}}

\bigskip

\setlength{\tabcolsep}{2pt}
\begin{tabular}{cccccccccccccccc}
\tt A & \tt C & \tt T & \tt T & \tt A & \tt G & \tt C & \dots & \tt C & \tt T & \tt A & \tt A & \tt C & \tt G & \tt \$ \\
0 & 1 & 2 & 3 & 4 & 5 & 6 & & 1121 & 1122 & 1123 & 1124 & 1125 & 1126 & 1127
\end{tabular}
\setlength{\tabcolsep}{6pt}

\bigskip

\resizebox{\textwidth}{!}
{\setlength{\tabcolsep}{2pt}
\begin{tabular}{rcl}
$B$ & $=$ &
00000000000000000000000000000000000000000000000000000000000000000000000000000000000000000000000000010\\
&&
00000000000000000000000000000001000000000000000000000000001000000000000000000000000000010000000000000 \dots \\
&& 
00000000000000000100000000000000000000000000000000000000000000000000000000100000000000000000000000000 \\
&&
00000010000000000000000000000000000000000000000000000000000000000000000000000000000000000000000000000
\end{tabular}
\setlength{\tabcolsep}{6pt}}

\bigskip
\bigskip

\begin{lrbox}{\myv}\begin{minipage}{1.125\textwidth}
\begin{tabular}{c@{\hspace{5ex}}c}
\begin{tabular}{r|rccl}
$i$ & $\SA [i]$ & $\LCP [i]$ & $\BWT [i]$ & context\\
\hline
0 & 303 & 0 & \verb!$! & \\
1 & 302 & 0 & \verb!5! & \verb!$! \\
2 & 102 & 1 & \verb!5! & \verb!$.__CXUGC<!\dots \\
3 & 210 & 1 & \verb!5! & \verb!$<.__\<J\N!\dots \\
4 & 156 & 2 & \verb!5! & \verb!$<<J__.\GG!\dots \\
5 & 174 & 8 & \verb!5! & \verb!$<<J__.\\\!\dots \\
\vdots & \vdots & \vdots & \vdots & \vdots
\end{tabular}
&
\begin{tabular}{r|rccl}
$i$ & $\SA [i]$ & $\LCP [i]$ & $\BWT [i]$ & context\\
\hline
\vdots & \vdots & \vdots & \vdots & \vdots \\
298 & 15 & 4 & \verb!'! & \verb!cKNJX5$=c<!\dots \\
299 & 268 & 1 & \verb!X! & \verb!cN/A\\\GQ2!\dots \\
300 & 230 & 2 & \verb!X! & \verb!cN2C<\\\G@!\dots \\
301 & 286 & 2 & \verb!X! & \verb!cNJ_\\\GQ2!\dots \\
302 & 249 & 2 & \verb!X! & \verb!cNQC<\\\GQ!\dots \\
303 & 67 & 1 & \verb!=! & \verb!c_cA\2X\G3!\dots
\end{tabular}
\end{tabular}
\end{minipage}\end{lrbox}

\resizebox{\textwidth}{!}{\usebox\myv}

\bigskip

\setlength{\tabcolsep}{2pt}
\begin{tabular}{cccccccccccccccc}
\tt = & \tt c & \tt < & \tt J & \tt \_ & \tt c & \tt A & \dots & \tt @ & \tt ' & \tt + & \tt 9 & \tt 2 & \tt 5 & \tt \$ \\
0 & 1 & 2 & 3 & 4 & 5 & 6 & & 296 & 297 & 298 & 299 & 300 & 301 & 302
\end{tabular}
\setlength{\tabcolsep}{6pt}

\bigskip

\setlength{\tabcolsep}{2pt}
\begin{tabular}{rcl}
$B$ & $=$ &
0000000000000000000001000000000000000000 \dots\\
&& 0001000000000000000000000000000000000001
\end{tabular}
\setlength{\tabcolsep}{6pt}

\bigskip
\bigskip

\begin{lrbox}{\myv}\begin{minipage}{1.125\textwidth}
\begin{tabular}{c@{\hspace{5ex}}c}
\begin{tabular}{r|rccl}
$i$ & $\SA [i]$ & $\LCP [i]$ & $\BWT [i]$ & context\\
\hline
0 & 236 & 0 & \tt \verb!$! & \\
1 & 38 & 0 & \tt \verb!<! & \tt \verb!#'C6J$=c_c!\dots \\
2 & 137 & 3 & \tt \verb!2! & \tt \verb!#'CN$NN__\!\dots \\
3 & 211 & 2 & \tt \verb!\! & \tt \verb!#'_N\5$XcN!\dots \\
4 & 185 & 1 & \tt \verb!2! & \tt \verb!#+J<5$NQC<!\dots \\
5 & 82 & 1 & \tt \verb!N! & \tt \verb!#.'$_c=\2X!\dots \\
\vdots & \vdots & \vdots & \vdots & \vdots
\end{tabular}
&
\begin{tabular}{r|rccl}
$i$ & $\SA [i]$ & $\LCP [i]$ & $\BWT [i]$ & context\\
\hline
\vdots & \vdots & \vdots & \vdots & \vdots \\
231 & 5 & 2 & \tt \verb!_! & \tt \verb!cA\2X<G2@'!\dots \\
232 & 29 & 1 & \tt \verb!'! & \tt \verb!cKNJ<5$c<#!\dots \\
233 & 15 & 4 & \tt \verb!'! & \tt \verb!cKNJX5$X\G!\dots \\
234 & 175 & 1 & \tt \verb!X! & \tt \verb!cN2C<\#G@2!\dots \\
235 & 219 & 2 & \tt \verb!X! & \tt \verb!cNJ_\\\GQ2!\dots \\
236 & 45 & 1 & \tt \verb!=! & \tt \verb!c_cA#G3K@$!\dots
\end{tabular}
\end{tabular}
\end{minipage}\end{lrbox}

\resizebox{\textwidth}{!}{\usebox\myv}

\bigskip

\setlength{\tabcolsep}{2pt}
\begin{tabular}{cccccccccccccccc}
\tt = & \tt c & \tt < & \tt J & \tt \_ & \tt c & \tt A & \dots & \tt @ & \tt ' & \tt + & \tt 9 & \tt 2 & \tt 5 & \tt \$ \\
0 & 1 & 2 & 3 & 4 & 5 & 6 & & 229 & 230 & 231 & 232 & 233 & 234 & 235
\end{tabular}
\setlength{\tabcolsep}{6pt}

\bigskip

\setlength{\tabcolsep}{2pt}
\begin{tabular}{rcl}
$B$ & $=$ &
0000000000000000000001000000000000010000 \dots \\
&&
0000000000100000000001000000000000000001
\end{tabular}
\setlength{\tabcolsep}{6pt}

\caption{The first and last lines of the augmented FM-indexes for the KATKA kernel in Figure~\ref{fig:kernel} {\bf (top)} and the minimizer digest in Figure~\ref{fig:digests} {\bf (bottom)}.}
\label{fig:indexes}
\end{center}
\end{figure}

Consider the pattern $P = \mathtt{GGATGGGCTAGACGATCTTCTGTG}$, which we obtained by choosing the substring {\tt GGGTGGGTTAGACGATCTTCTGTA} of toy genome 9 in Figure~\ref{fig:genomes} (numbering the genomes from 0) and changing two characters.  The MEM table of $P$ with respect to all the toy genomes is shown on the left in Figure~\ref{fig:MEM_tables}.  The MEM table of $P$ with respect to the 4th-order KATKA kernel with {\tt \$}s and {\tt \#}s shown in Figure~\ref{fig:kernel}, is shown in the center of Figure~\ref{fig:MEM_tables}.  (The MEM table of $P$ with respect to the 5th-order KATKA kernel is the same as its MEM table with respect to the genomes.)  The minimizer digest of $P$ with $w = 10$ is {\tt Q.} and the MEM table of that with respect to the minimizer digest of the collection is shown on the right of Figure~\ref{fig:MEM_tables}; the MEM table with respect to the 2nd-order KATKA kernel of the minimizer digest is the same as the MEM table with respect to the minimizer digest.

\begin{figure}[t]
\begin{center}
\begin{tabular}{ccc}
\begin{tabular}{rrr}
MEM & first & last \\
\hline
\tt GGATGGGCTAG & 13 & 13 \\ 
\tt TAGACGATCTTCTGT & 9 & 9 \\
\tt TGTG & 0 & 1 \\
&& \\
&& \\
&& \\
&& \\
&& \\
&& \\
&&
\end{tabular}
&
\begin{tabular}{rrr}
MEM & first & last \\
\hline
\tt GGATGGG & 13 &13 \\
\tt GGGC & 6 & 15 \\
\tt GGCT & 12 & 14 \\
\tt GCTAG & 15 & 15 \\
\tt TAGA & 0 & 15 \\
\tt AGACG & 15 & 15 \\
\tt GACGATC & 11 & 11 \\
\tt ATCTTCT & 0 & 15 \\
\tt TCTGT & 8 &8 \\
\tt TGTG & 0 & 1
\end{tabular}
&
\begin{tabular}{rrr}
MEM & first & last \\
\hline
\tt Q & 8 & 15 \\ 
\tt . & 4 & 11 \\
&& \\
&& \\
&& \\
&& \\
&& \\
&& \\
&& \\
&&
\end{tabular}
\end{tabular}
\caption{The MEM tables of $P$ with respect to the toy genomes in Figure~\ref{fig:genomes} (left), the 4th-order KATKA kernel in Figure~\ref{fig:kernel} (center), and the minimizer digests in Figure~\ref{fig:digests} (right).}
\label{fig:MEM_tables}
\end{center}
\end{figure}

Since $P$ comes from toy genome 9, following Wood, Lu and Langmead's~\cite{wood2019improved} terminology in their presentation of Kraken~2, we classify MEMs' [first, last] ranges as {\em true positives} if they are exactly [9,9], {\em false positives} if they exclude 9 but are not empty, {\em vague positives} if they include 9 and at least one other number, and false negatives if they are empty.  The classification of the MEMs' ranges in Figure~\ref{fig:MEM_tables} are shown below:
\begin{center}
\begin{tabular}{cccc}
true positives & false positives & vague positives & false negatives\\
\hline
[9] & [0, 1], [8], [11] & [0, 15], [4, 11] & \\
    & [12, 14], [13], [15] & [6, 15], [8, 15] &
\end{tabular}
\end{center}
Notice the ranges for MEMs with respect to the toy genomes and the 4th-order KATKA kernel can never be empty (assuming every distinct character in $P$ occurs in the genomes at least once), so those ranges cannot be false negatives.  On the other hand, if we generate $P$ by changing characters in a way that disrupts every previous minimizer and creates new minimizers that are not in the minimizer digest of the genomes, then we can get MEMs with respect to the minimizer digest or to the 2nd-order KATKA kernel of the minimizer digest, whose ranges are empty.

Looking at the MEM table of $P$ with respect to the toy genomes, it is intuitive to give more weight to the longer MEM, which occurs only in genome 9.  If on this basis we guess correctly that $P$ came from genome 9, then we can consider $P$ a true positive with respect to the toy genomes; unfortunately, the same is not true with respect to the 4th-order KATKA kernel, nor to the minimizer digest with $w = 10$.

\section{Experiments}
\label{sec:experiments}

In order to present a concise comparison of results obtained with a full dataset with those obtained with KATKA kernels, minimizer digests, and KATKA kernels of minimizer digests, for this section we focus on true-positive rates rather than whole MEM tables.  We classify a read as a true positive if its longest MEM is a true positive (or all its longest MEMs, in the case of a tie).

We wrote the code for our experiments (which computes full MEM tables) in C++ using SDSL~\cite{sdsl} and posted it at \url{https://github.com/draessld/KATKA2}.  We ran our experiments on a server at the Department of Computer Science of the Czech Technical University in Prague with 128 AMD EPYC 7742 64-Core CPUs and 504 GiB of RAM, running GNU/Linux Kernel 5.15.0.

We chose 1000 bacterial genera consecutive in the phylogenetic tree for 138.1 release of the SILVA SSU Ref NR99 database~\cite{silva} of ribosomal RNA (rRNA) gene sequences.  We concatenated the gene sequences for the genera, separated by {\tt \$}s, and built augmented FM-indexes for that 167328343-character concatenation, and KATKA kernels, minimizer digests, and KATKA kernels of minimizer digests for it with various parameter settings:
\begin{itemize}
    \item for KATKA kernels of the original concatenation, we used $k = 5, 10, 15, 20, \ldots, 45, 50, 100$;
    \item for minimizer digests, we used 3-mers as minimizers and set $w = 5, 10, 15, 20, \ldots, 45, 50$;
    \item for KATKA kernels of the minimizer digests, we used $k = 5, 10, 15, 20, \ldots, 45, 50$ and the same $w$ values.
\end{itemize}
We included the kernel with $k = 100$ of the original concatenation to show that as $k$ increases, the true-positive rate does approach the rate achieved with an index of the original concatenation.

\begin{figure}[t]
    \includegraphics[width=\textwidth]{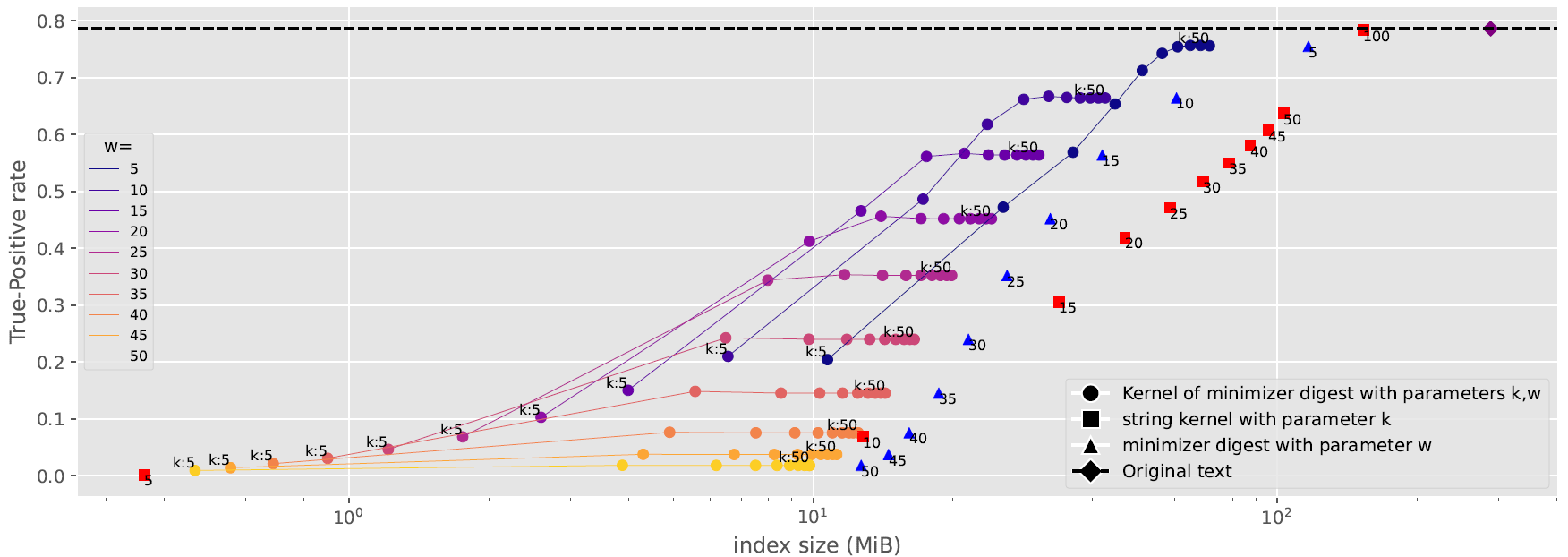}
    \caption{The index size in MiB and the true-positive rate as a percentage, for the original dataset and various KATKA kernels, minimizer digests, and KATKA kernels of minimizer digests.}
    \label{fig:results}
\end{figure}

\begin{figure}[t]
    \includegraphics[width=\textwidth]{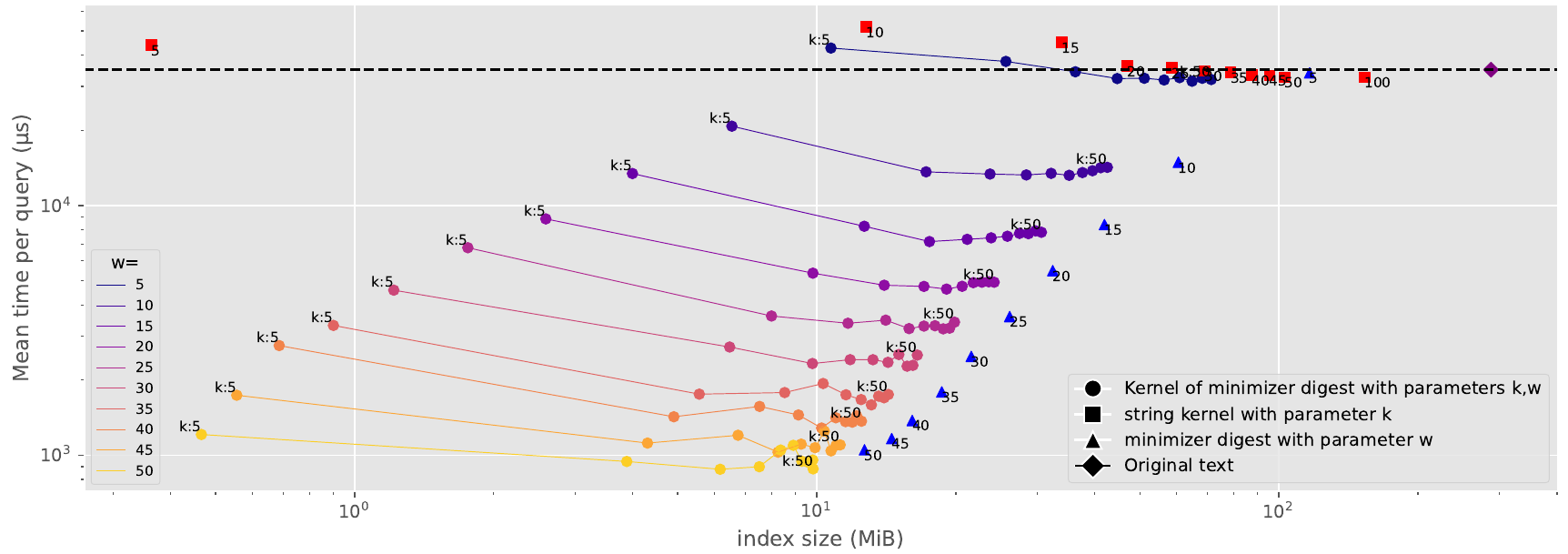}
    \caption{The index size in MiB and the mean search time per read in microseconds, for the original dataset and various KATKA kernels, minimizer digests, and KATKA kernels of minimizer digests.}
    \label{fig:results_times}
\end{figure}

For each genus $g$, we simulated 500 reads of 200 base pairs each by choosing a random starting location in the reference sequence for $g$ and mutating 1\% percent of the bases uniformly across the read to simulate sequencing error.  For each read and each index, we found all the read's longest MEMs and checked whether all their [first, last] ranges contained only the ID of the reference sequence for $g$.  Figure~\ref{fig:results} shows the index sizes and true-positive rates over all 500\,000 simulated reads, and Figure~\ref{fig:results_times} shows an average search time per read.

Clearly, we can achieve significant compression while only slightly decreasing the true-positive rate and without paying a penalty in search time, especially with KATKA kernels of minimizer digests: for example, with $k = 30$ and $w = 5$ our index took 56.5 MiB and achieved a true-positive rate of 74.3\%, compared to 287.9 MiB and 78.6\% with an index for the full dataset; our index is also slightly faster than the index for the full dataset.  This is a better compression/accuracy/speed tradeoff than we achieve with either kernelization or minimizers alone.

If we are willing to sacrifice the true-positive rate moderately, we can increase $w$ and also achieve significant speedups.  FM-indexes over minimizer digests are known to be usually significantly faster than indexes over the original datasets, both because some characters are not represented in the digests and because we use a backward step for each minimizer rather than for each character, incurring fewer cache misses.  Interestingly, however, we achieved slightly better speedups with both kernelization and minimizers than with either separately.

\section{Conclusions and future work}
\label{sec:conclusion}

Figure~\ref{fig:results} strongly confirms our conjecture from Subsection~\ref{subsec:kernels_of_digests} that KATKA kernels of minimizer digests can inherit the strengths of both.  In the near future we plan to experiment also with varying the width of minimizers (for simplicity, in this paper we always used 3-mers).  Later, we plan to incorporate indexing KATKA kernels of minimizer digests to build MEM tables --- with more sophisticated classifications that take advantage of all the information in those tables --- into a full pipeline for taxonomic classification of reads.

The confirmation of our conjecture may be useful for other applications as well, when we are dealing with repetitive datasets and want the flexibility of an augmented FM-index (instead of an $r$-index or a grammar-based index, for example) but kernelization has still had less impact than we might have hoped, because setting the parameter $k$ high enough to allow for the pattern lengths used in practice results in poor compression.  For example, an obvious question that arises from our work is whether Valenzuela et al.'s~\cite{PanVC} PanVC tool can achieve interesting tradeoffs between compression and accuracy using kernelization of minimizer digests, instead of only kernelization.

\section*{Acknowledgments}

Many thanks to Sana Kashgouli and Finlay Maguire for helpful discussions.
D.D.\ funded by CTU grant No. SGS23/205/OHK3/3T/18 and by ROBOPROX reg. no. CZ.02.01.01/00/22\_\allowbreak 008/0004590.
O.Y.A.~and B.L.~funded by NIH grants R35GM139602 \& R01HG011392 and NSF grant DBI-2029552.
T.G.\ funded by NSERC Discovery Grant RGPIN-07185-2020.
G.M.\ funded by the Italian Ministry of Health, POS 2014-2020, project ID T4-AN-07, CUP I53C22001300001, by INdAM-GNCS Project CUP E53C23001670001 and by PNRR ECS00000017 Tuscany Health Ecosystem, Spoke~6 CUP I53C22000780001.
G.N.\ funded by Basal Funds FB0001, ANID, Chile.

\end{document}